\documentclass[twocolumn]{aastex62}
\usepackage{amsmath,amstext}
\usepackage{units}
\usepackage{url}
\usepackage{xspace}

\newcommand{\E}[1]{\times10^{#1}}
\newcommand{\msol}{ \, M_\odot}
\newcommand{\smpy}{ \, M_\odot \, {\rm yr^{-1}}}

\newcommand{\bi}{\begin{itemize}}
\newcommand{\ei}{\end{itemize}}
\newcommand{\commentOut}[1]{}

\newcommand{\mesa}{\texttt{MESA}\xspace}

\shortauthors{Shen \& Quataert}

\begin{document}


\title{\bf \Large{Binary interaction dominates mass ejection in classical novae}}

\author[0000-0002-9632-6106]{Ken J.\ Shen}
\affiliation{Department of Astronomy and Theoretical Astrophysics Center, University of California, Berkeley, CA 94720, USA}

\author[0000-0001-9185-5044]{Eliot Quataert}
\affiliation{Department of Astrophysical Sciences, Princeton University, Princeton, NJ 08544, USA}


\begin{abstract}

Recent observations suggest our understanding of mass loss in classical novae is incomplete, motivating a new theoretical examination of the physical processes responsible for nova mass ejection.  In this paper, we perform hydrodynamical simulations of classical nova outflows using the stellar evolution code \mesa.  We find that, when the binary companion is neglected, white dwarfs with masses $\gtrsim 0.8 \msol$ successfully launch radiation-pressure-driven optically thick winds that carry away most of the envelope.  However, for most of the mass loss phase, these winds are accelerated at radii beyond the white dwarf's Roche radius assuming a typical cataclysmic variable donor.  This means that, before  a standard optically thick wind can be formed, mass loss  will instead be initiated and shaped by the binary interaction.  An isotropic optically thick wind is only successfully launched when the acceleration region recedes within the white dwarf's Roche radius, which occurs after most of the envelope has already been ejected.  The interaction between these two modes of outflow -- a first phase of slow, binary-driven,  equatorially focused mass loss  encompassing most of the mass ejection and a second phase consisting of a fast, isotropic, optically thick wind -- is consistent with observations of aspherical ejecta and signatures of multiple outflow components.  We also find that isolated lower-mass white dwarfs $\lesssim 0.8 \msol$ do not develop unbound optically thick winds at any stage, making it even more crucial to consider the effects of the binary companion on the resulting outburst.  

\end{abstract}



\section{Introduction and the physics of classical nova outbursts}

A classical nova outburst results from the ignition of a hydrogen-rich envelope on a white dwarf (WD) accreted from a binary companion in a cataclysmic variable (CV) system (see \citealt{bode08a} and \citealt{chom21a} for reviews).\footnote{Recurrent novae are novae with short enough recurrence times that multiple outbursts have been observed in human history.  Novae in symbiotic systems arise from the same basic physics, but they occur in much more widely separated binaries with giant donors.}  Once a sufficient envelope mass has been accumulated ($\sim 10^{-4}-10^{-6} \msol$, depending on properties such as the accretion rate and the WD mass and temperature; \citealt{pk95,tb04,yaro05}), thermonuclear fusion of the hydrogen-rich fuel generates energy faster than it can diffuse away, and a convective burning shell is born.

This convective zone heats on a  timescale given by the ratio of the envelope's internal energy to its nuclear luminosity:
\begin{align}
	t_{\rm heat} =& \ \frac{ \int_{M_{\rm core}}^{M_{\rm total}} c_P T dm }{ \int_{M_{\rm core}}^{M_{\rm total}} \epsilon dm} \sim f  \frac{c_{P,{\rm base}} T_{\rm base}}{\epsilon_{\rm base}} \nonumber \\
	 \sim & \  \unit[10]{yr}  \left( \frac{f}{4} \right) \left( \frac{0.6}{X} \right) \left( \frac{0.2}{Z_{\rm CNO}} \right)  \nonumber \\
	& \times \left( \frac{\unit[3\E{3}]{g \, cm^{-3}}}{\rho_{\rm base}} \right) \left( \frac{T_{\rm base}}{\unit[3\E{7}]{K}} \right)^{5/3} \nonumber \\
	& \times \exp \left[ \frac{49.008 }{(T_{\rm base}/ \unit[3\times10^7]{K})^{1/3}} - 49.008 \right] ,
\end{align}
where the total mass, $M_{\rm total}$, is the sum of the underlying core mass, $M_{\rm core}$, and the outbursting envelope mass, $M_{\rm env}$.  The factor of $f \sim4$  is estimated from simulations described later in this paper and approximately accounts for the extent of the burning region being smaller than the region encompassing the bulk of the internal energy in the convective zone. The subscript ``base'' refers to quantities evaluated at the base of the envelope, the specific heat at constant pressure, $c_P$, is assumed to be the ideal gas value, $5 k / 2 \mu m_p$ (as roughly appropriate at the onset of convection), and the energy generation rate, $\epsilon$, is evaluated in the cold CNO limit, where the $^{14}$N$+p$ reaction is the rate-limiting step \citep{lemu06}.  The fiducial values of the hydrogen mass fraction, $X$, and the CNO mass fraction, $Z_{\rm CNO}$, are representative of a C/O-enriched envelope, as commonly observed in nova ejecta \citep{gehr98}.

\begin{figure}
  \centering
  \includegraphics[width=\columnwidth]{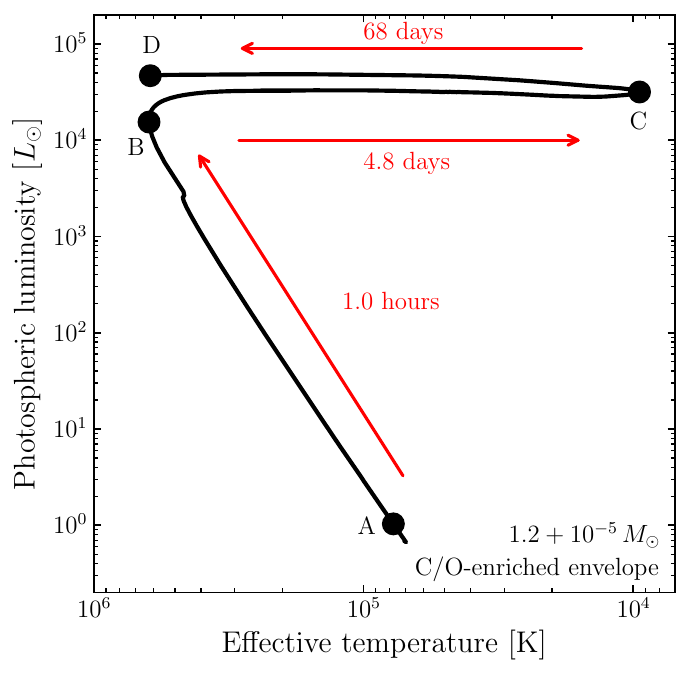}
	\caption{Photospheric luminosity vs.\ effective temperature for a  simulation of a $1.2 \msol$ WD with a $10^{-5} \msol$ C/O-enriched outbursting nova envelope.  The model is evolved with \mesa and does not include any effects of a companion star.  See Sections \ref{sec:desc} and \ref{sec:results} for more details.}
	\label{fig:lvsteff}
\end{figure}

The heating convective zone  grows outwards in mass (and perhaps inwards as well if convective dredge-up and overshooting are efficient).  As the outer edge of the convective zone nears the photosphere, radiation is able to diffuse out of the star, the photospheric luminosity increases dramatically, and the WD climbs back up the WD cooling sequence with a roughly constant radius.  Figure \ref{fig:lvsteff} shows the evolution in the Hertzsprung-Russell diagram of a  model of a $1.2 \msol$ WD with a $10^{-5} \msol$  outbursting envelope that is enriched with 10\% $^{12}$C and 10\% $^{16}$O by mass (see Sections \ref{sec:desc} and \ref{sec:results} for simulation details).  This initial phase of rising luminosity between point A, where the photospheric luminosity is $1 \, L_\odot$, and point B, the time of maximum effective temperature, $T_{\rm eff}$, happens very rapidly for this case, with the luminosity increasing by four orders of magnitude in just one hour.

This brightening continues until the radiative luminosity nears the Eddington limit,
\begin{align}
	L_{\rm Edd} = & \ \frac{4 \pi G M_{\rm core} c }{ \kappa } \nonumber \\
	= & \ 1.7\E{38} \, {\rm erg \, s^{-1}} \left( \frac{M_{\rm core}}{M_\odot} \right)  \left( \frac{\unit[0.3]{cm^2 \, g^{-1}}}{\kappa} \right),
\end{align}
where the mass of the WD core dominates the total mass of the star, and $\kappa$ is the opacity.  After this point, the envelope expands at a roughly constant luminosity, causing the effective temperature to decrease from its maximum value.  The brief time spent near the maximum effective temperature leads to a short-lived ultraviolet or X-ray nova precursor, depending on the WD mass, as calculated by \cite{hill14a} and observed in the ultraviolet by \cite{cao12a} and in the X-rays by \cite{koen22a}.  It is important to note that initially, this phase of expansion is relatively gentle and essentially hydrostatic for most of the envelope's mass, with expansion velocities that are slower than the local speed of sound.

For cases where  the convective turnover timescale is short enough, convection will transport $\beta$-unstable nuclei ($^{13}$N, $^{14}$O, $^{15}$O, and $^{17}$F) from CNO-burning at the base of the envelope   up to the outer layers before they have had a chance to decay.  If there is significant pollution of the envelope by core material rich in carbon and oxygen, the energy generation rate from these decays will be sufficient to directly eject material from the outer layers at high velocities $> \unit[1000]{km \, s^{-1}}$(\citealt{star16a} and references therein);   for the model shown in Figure \ref{fig:lvsteff}, no material is lost  in this way.  We discuss this  ejection mechanism further in Section \ref{sec:results}, but we note here that when  it  does occur,  it only unbinds a small amount of material.  The vast majority of the envelope mass during this stage expands slowly and remains bound to the WD.

The rest of the nova envelope, minus any small mass unbound by the surface $\beta$-decays, continues to expand until one of two things happens: either the envelope overflows the WD's Roche lobe and encounters the companion's gravitational potential, or an optically thick wind is launched.  Most previous theoretical studies of classical nova outbursts have chosen to neglect the binary companion and have focused on the second outcome.  We now briefly review the sequence of events that arises if the companion is neglected before highlighting observational challenges to this picture.

Once the envelope has expanded sufficiently, part of it will traverse a peak in the opacity at a temperature of  $\sim \unit[1.5\E{5}]{K}$ due to bound-bound transitions of iron-group elements \citep{igle90a}.  This ``iron opacity bump'' results in a localized trough in the Eddington luminosity, causing the luminosity to become super-Eddington even though its absolute value stays constant.  Radiative diffusion is no longer sufficient to carry the luminosity, so convection sets in.  However, the inferred convective velocity from mixing-length theory required to carry this high flux is supersonic, and thus convection is insufficient to transport the luminosity.  Instead, an optically thick wind is formed, which drives a sustained outflow that ejects most of the nova envelope.  

During this phase, the photospheric radius first increases until a minimum effective temperature is reached (corresponding to the 5-day evolution between points B and C for the model shown in Fig.\ \ref{fig:lvsteff}), and then, as envelope mass is lost, the photosphere recedes.  This phase is the longest of the outburst, taking $\unit[68]{d}$ for the model in Figure \ref{fig:lvsteff} to evolve between points C and D.

Once the envelope mass has decreased to the maximal envelope mass that can support a thin, steadily burning, thermally stable hydrostatic solution \citep{fuji82b,nomo07,sb07,wolf13a}, mass loss ends.  The remaining bound envelope continues to fuse hydrogen at its base, emitting as a supersoft X-ray source due to the contraction of the photosphere.  However, the actual appearance of the nova system may not match the WD's emission due to obscuration by the previous phase of mass ejection.  Eventually, the consumption of fuel reduces the envelope mass below the minimum stable solution, burning ceases, and the WD evolves back down the cooling sequence.  Continued accretion provides a fresh source of hydrogen-rich fuel until enough has accumulated to trigger the next nova outburst, and the cycle begins again.

Having reviewed the standard picture of mass loss due to near-Eddington winds, we now summarize  observational evidence that has mounted against the main assumption of a spherically symmetric, optically thick wind being the main driver of nova mass loss  \citep{chom21a}.  Resolved imaging of novae during eruption \citep{chom14a} and years and decades afterwards (e.g.,  \citealt{slav95a,soko17a}) shows clear evidence for aspherical ejecta and multiple modes of mass ejection.  This is also seen in  spectral indicators (e.g., \citealt{aydi20b}) and the detection  of shock interaction in many nearby classical novae \citep{acke14a,west16a,finz18a,nels19b,aydi20a,soko20a}.  Optically thick winds also fail to explain the super-Eddington luminosities (e.g., \citealt{schw01a,aydi18a}) and the wide variety of behaviors \citep{stro10a} demonstrated by some nova light curves.

In this study, we perform hydrodynamical simulations of classical nova outbursts on isolated WDs using the stellar evolution code \mesa \citep{paxt11,paxt13,paxt15a,paxt18a}.    Our results confirm the previously described sequence of events for the launching of optically thick winds for WD masses $\gtrsim 0.8 \msol$.  However, for lower-mass WDs,  none of the models we run achieves a successful wind solution; instead, the envelope  remains bound for its entire evolution as the wind is either choked or is never launched at all. 

Crucially, we find that, even for successful winds, the velocity at the location of the WD's Roche radius is slower than the typical orbital velocities of WDs in CVs for most of the mass loss phase, because the acceleration region is outside of the WD's Roche lobe.  As a result, standard optically thick winds cannot form, and  binary interaction will  instead initiate the majority of mass loss, focusing it towards the equatorial plane.  It is only during the final stages of the nova outburst, when the acceleration region recedes within the WD's Roche lobe, that a spherical optically thick wind can form.  The interaction of this fast, isotropic wind with the  prior phase of slow, equatorial outflow is likely responsible for the previously outlined observations of aspherical, self-interacting ejecta.

There have been numerous \mesa studies of various properties of classical nova systems  (e.g., \citealt{deni13a,deni14a,wolf13a,wolf18b,chen19a,ginz21a,leun21b,guo22a}), but the present work is the first to use  it to study hydrodynamical nova outflows.   We outline our simulations in Section \ref{sec:desc} and describe our results in Section \ref{sec:results}.  We compare to previous studies of nova outbursts in Section \ref{sec:comp}.  We summarize our findings and discuss their broader implications in Section \ref{sec:conc}.


\section{Description of simulations}
\label{sec:desc}

We use \mesa\footnote{http://mesa.sourceforge.net, version 10398; default options used unless otherwise noted.} to evolve nova outbursts, assuming no interaction with a companion star, for  a range of WD core masses ($0.6$, $0.8$, $1.0$, and $1.2 \msol$) and envelope masses that bracket  the expected ignition masses found by evolutionary calculations \citep{pk95,tb04,yaro05,deni13a}.  We ignore the pre-outburst accretion history and instead take the mass coordinate of the base of the convective zone during the outburst as a given.  We thus avoid the difficulty of setting up the correct equilibrium conditions of the underlying core, which depend on many factors including the ages of the WD and CV, and which otherwise play an important role in determining the ignition mass of the  outbursting envelope  \citep{tb04,epel07}. 

We are also agnostic to the details of the envelope's enrichment with core material, which is frequently observed in nova ejecta \citep{gehr98} and may occur a variety of ways, including chemical diffusion \citep{pk84,pk95,kp85,yaro05}, rotationally induced shear mixing \citep{kt78,lt87,ks89,alex04}, and convectively induced shear mixing and overshooting  \citep{woos86,gl95,glt97,glas12a,rosn01,casa10a,casa11a,casa11b,casa16a,casa18a}.  As with the envelope mass, we take the envelope composition as a given  because of uncertainties in the strengths of the various enrichment mechanisms and explore  two compositions: solar composition \citep{aspl09a}, which is equivalent to the assumption that no mixing occurs with the underlying C/O core, and an enhanced C/O composition, with 80\% solar composition by mass plus an additional 10\% by mass of $^{12}$C and $^{16}$O each.  The composition of the accreted material may not be solar, which can induce changes to the ignition mass \citep{pier00,star00,jh07,sb09a,chen19a,kemp22a,hill22a}, but since we parametrize the convective envelope masses by hand in this study, we bypass this complication.

We construct our WD models by first creating pure C/O objects with specified masses between $0.6$ and $ 1.2 \msol$ with  nuclear burning disabled.  To prevent spurious numerical mixing between the envelope and the core during the outburst, which would further enrich the envelope with C/O, we accrete a $10^{-4} \msol $ pure $^4$He buffer before adding the envelope with the desired composition and mass.  We specify the mass fractions of  $^1$H, $^4$He, $^{12}$C, $^{14}$N, and $^{16}$O in the envelope material.  The entire star is then allowed to cool until the center reaches $ \unit[2\E{7}]{K}$, and nuclear burning is turned back on.   We use the  \texttt{cno\_extras} burning network, which includes $^1$H, $^{3-4}$He, $^{12-13}$C, $^{13-15}$N, $^{14-18}$O, $^{17-19}$F, $^{18-20}$Ne,  and $^{22,24}$Mg, and the reactions linking them with rates from the JINA Reaclib database \citep{cybu10a}.  

Due to the low temperatures resulting from cooling the entire star, most models do not experience a convective thermonuclear runaway when nuclear burning is turned on.  We thus input extra heating at the base of the envelope until the convective burning is self-sustained, at which point the extra heating is turned off.    Once convective burning begins, we enable a variety of flags in \mesa to follow the hydrodynamic evolution of the outburst.    Further details regarding this implementation and the inlists used for the calculations are located in the Appendix.

Simulations that result in successful optically thick winds are evolved until the mass outflow  begins to cease, at which point numerical issues lead to small timesteps and the simulations are ended.  Models that do not launch winds enter a long-lived steady state and are halted after several weeks of computational time.


\section{Simulation results}
\label{sec:results}


\subsection{Models with successful winds}

\begin{figure}
  \centering
  \includegraphics[width=\columnwidth]{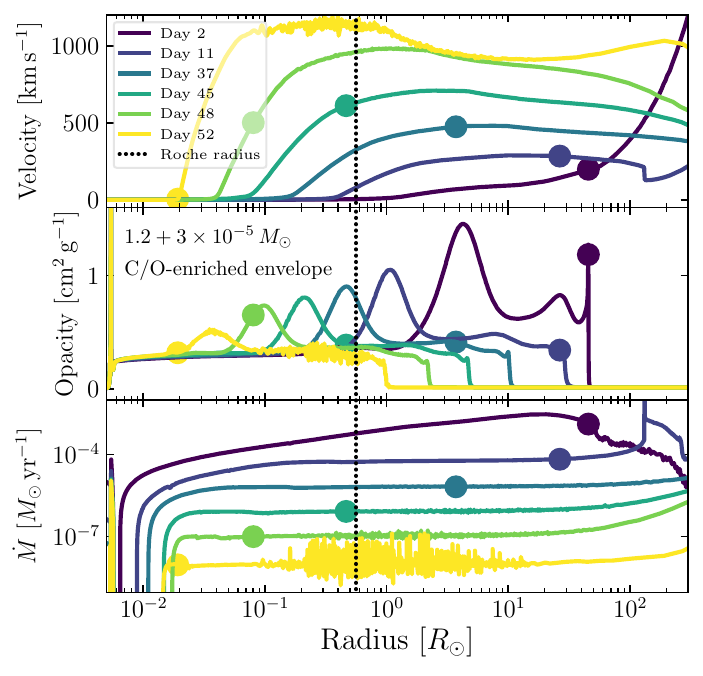}
	\caption{Velocity (top panel), opacity (middle panel), and mass flux (bottom panel) profiles for the $1.2 + 3\E{-5} \msol$ C/O-enriched envelope  calculation.  The location of the $\tau=2/3$ photosphere is shown as a circle for each profile, and the location of the Roche radius assuming a $0.3 \msol$, $0.3 \, R_\odot$ Roche-lobe-filling donor is shown as a dotted line.  Times are given with respect to the time of maximum effective temperature.}
	\label{fig:prof1_12_3e-5_co}
\end{figure}

\begin{figure}
  \centering
  \includegraphics[width=\columnwidth]{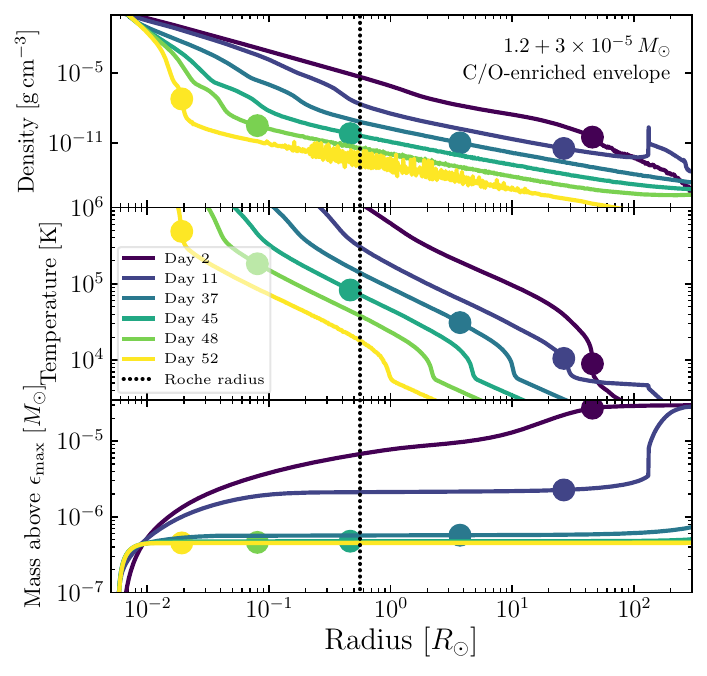}
	\caption{Same as Fig.\ \ref{fig:prof1_12_3e-5_co}, but showing radial profiles of the density (top panel), temperature (middle panel), and mass above the location of maximum energy generation (bottom panel).  Comparison of Figs.\ \ref{fig:prof1_12_3e-5_co} and \ref{fig:prof2_12_3e-5_co} shows that most of the envelope is lost between days 2 -- 11 when the outflow velocities at the Roche radius are low.  In reality, binary interactions thus dominate the properties of the ejected mass.}
	\label{fig:prof2_12_3e-5_co}
\end{figure}

In Figures \ref{fig:prof1_12_3e-5_co} and \ref{fig:prof2_12_3e-5_co}, we show radial profiles of various quantities of interest for the $1.2 \msol$ WD with a C/O-enriched envelope of $3\E{-5} \msol$.  The times of the profiles are given with respect to the time of maximum effective temperature.  The profile at day 2 shows evidence of high velocity ejection in the outer layers, with velocities surpassing $\unit[10^3]{km \, s^{-1}}$.  This is due to the convective mixing of  $\beta$-unstable nuclei (primarily $^{13}$N and $^{15}$O) into the surface layers of the star at a time when the surface has expanded enough for the decay energy to be sufficient to unbind material (see \citealt{star16a} and references therein).  However, such a phase only lasts a short duration before convection recedes from the surface layers.  As a result, only a very small amount of mass ($<10^{-7} \msol$) reaches velocities $>\unit[1000]{km \, s^{-1}}$ at the beginning of this simulation.   Furthermore, C/O enrichment is necessary to yield enough $\beta$-unstable nuclei and subsequent sufficient specific energy release to power this initial outflow; the $1.2 \msol$ WD with a solar composition $3\E{-5} \msol$ envelope does not exhibit this outer layer of fast-moving material.  The only other simulations that result in an initial phase of rapidly moving surface layers are the most massive C/O-enriched envelopes for each WD mass, but, as for the $1.2 + 3\E{-5} \msol$ model, in each case, $<10^{-7} \msol$ reaches velocities of $\unit[1000]{km \, s^{-1}}$.  This initial phase of rapid mass ejection is thus more important for larger envelope masses and higher C/O enrichment.  Future work will examine these effects, but it appears to make a  very small contribution to the total mass lost during typical novae.

By day 11, a steady-state optically thick wind has been established, with a mass flux that is constant through most of the envelope.  The acceleration region coincides with the iron opacity bump.  While most of the envelope has adjusted itself to be somewhat sub-Eddington with respect to electron-scattering opacity, the increased opacity at the iron bump leads to a locally super-Eddington luminosity.  Convection is active in this region but is inefficient, and as a result, a wind is launched.

\begin{figure}
	\includegraphics[width=\columnwidth]{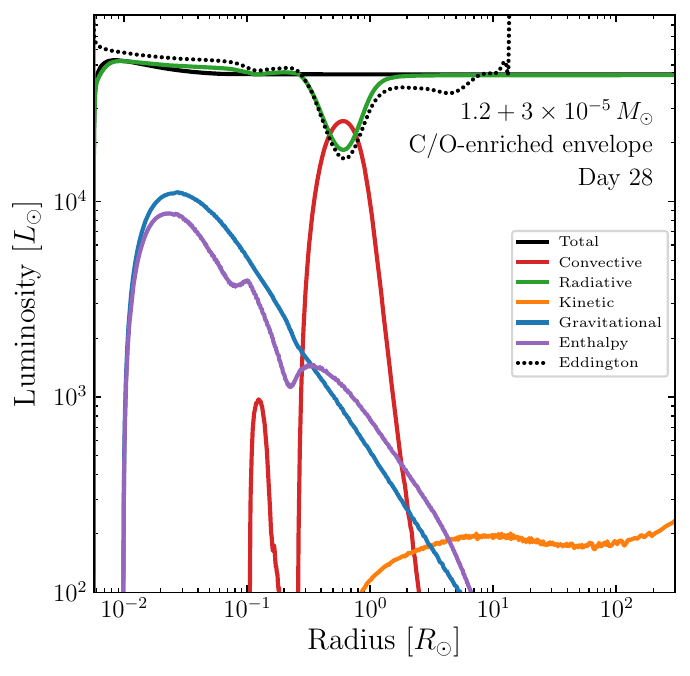}
	\caption{Radial profiles of various luminosities, as labeled, for the $1.2 + 3\E{-5} \msol$ C/O-enriched envelope model.}
	\label{fig:lumprof}
\end{figure}

Figure \ref{fig:lumprof} shows various luminosities within the $1.2+3\E{-5} \msol$, C/O-enriched envelope model at day 28, during this steady-state, optically thick wind phase.  The figure shows the kinetic energy luminosity, $\dot{M} v^2/2$ (orange), the power required to climb out of the potential well, $G M \dot{M} / r$ (blue), the enthalpy luminosity, $\dot{M} (u + P/\rho)$ (purple), where $u$ is the mass-specific energy, the radiative luminosity (green), and the convective luminosity (red).  The total luminosity, $\Lambda$ (black), is essentially constant throughout the envelope, as befits this steady-state solution, except for near the envelope base where the nuclear energy generation is located.  The iron opacity peak implies a trough in the Eddington luminosity (black dotted line), causing the luminosity to be mildly super-Eddington in that region.

At this time, the asymptotic kinetic power is a small fraction of the total luminosity; most of the nuclear burning luminosity emerges as radiative luminosity during this stage.  However, over the lifetime of the nova outburst, the majority of the energy release goes to overcoming the gravitational binding energy prior to and during the establishment of the steady-state wind: $E_{\rm bind} \gg E_{\rm lum}$, where the energy to unbind most of the nova envelope is
\begin{align}
	E_{\rm bind} \sim & \ \frac{G M_{\rm core} M_{\rm env}}{R_0} \nonumber \\
	\sim & \  \unit[5.3\E{46}]{erg} \left( \frac{M_{\rm core}}{M_\odot} \right) \left( \frac{M_{\rm env}}{10^{-4} \msol} \right) \nonumber \\
	& \times \left(  \frac{\unit[5\E{8}]{cm}}{R_0} \right) ,
\end{align}
and the total radiated luminosity is
\begin{align}
	E_{\rm lum} \sim & \ L_{\rm Edd} \, t_{\rm Edd} \nonumber \\
	\sim & \  \unit[4.3\E{44}]{erg} \left( \frac{M_{\rm core}}{M_\odot} \right) \left( \frac{\unit[0.3]{cm^2 \, g^{-1}}}{\kappa} \right) \nonumber \\
	&  \times \left( \frac{t_{\rm nova}}{\unit[1]{month}} \right) .
\end{align}
The initial WD radius is $R_0$, and the time the nova spends near the Eddington luminosity is $t_{\rm Edd}$.

\begin{figure}
  \centering
  \includegraphics[width=\columnwidth]{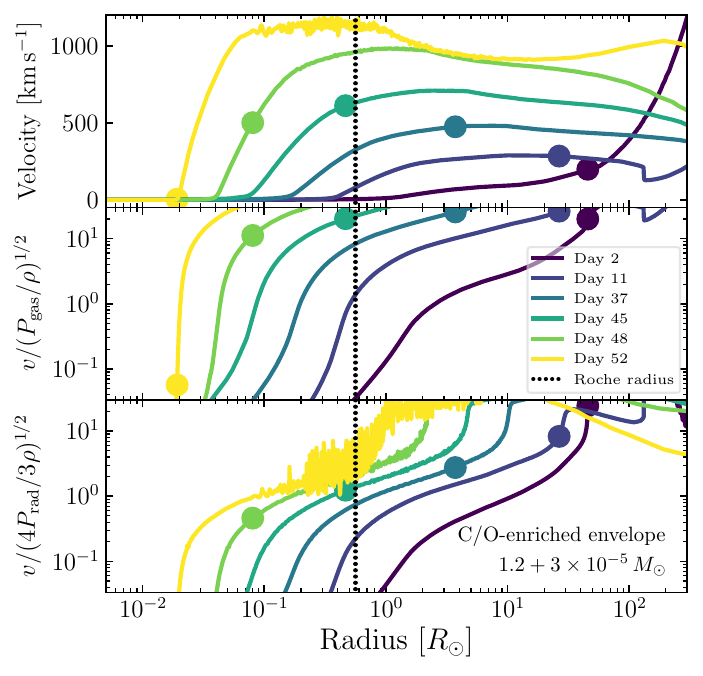}
	\caption{Same as Fig.\ \ref{fig:prof1_12_3e-5_co}, but showing radial profiles of the ratio of the velocity to the isothermal gas sound speed (middle panel) and the ratio of the velocity to the radiation sound speed (bottom panel).  The velocity profile is shown again in the top panel for clarity.}
	\label{fig:prof3_12_3e-5_co}
\end{figure}

Figure \ref{fig:prof3_12_3e-5_co} shows the ratio of the velocity to the isothermal gas sound speed (middle panel) and the radiation sound speed (bottom panel) at the same times as in Figures \ref{fig:prof1_12_3e-5_co} and \ref{fig:prof2_12_3e-5_co}.  As befits wind solutions, the velocities smoothly accelerate and cross a critical point where the velocity equals the isothermal gas sound speed \citep{bond52a,park58a,park65a,kh94}.  Crucially, up until   day 11, this critical point is outside of the WD's Roche radius (dotted lines in Figs.\ \ref{fig:prof1_12_3e-5_co}, \ref{fig:prof2_12_3e-5_co},  and  \ref{fig:prof3_12_3e-5_co}), assuming a $0.3 \msol$, $0.3 \, R_\odot$ Roche-lobe-filling companion, which we take as a typical donor in a classical nova system.\footnote{Most present-day CVs have periods below the period gap, implying donor masses $ \leq 0.2 \msol$ \citep{pala22a}.  However, this is a consequence of the deceleration in evolution as the donor mass decreases.  The rate of novae is highest when the donor mass is highest because the accretion rate is at its maximum and the ignition mass is at its minimum.  As a result, most of the novae during a CV's evolution will occur shortly after birth, when the donor mass is highest.  Thus, we use the mean of the pre-CV secondary mass distribution, $0.3 \msol$ \citep{zoro11a}, as our fiducial companion mass.}  Since the envelopes are dominated by radiation pressure, the radii at which they become truly hydrodynamic, with velocities exceeding the radiation sound speed, are even farther out in the envelope and remain outside the Roche radius until day 37.

\begin{figure}
	\includegraphics[width=\columnwidth]{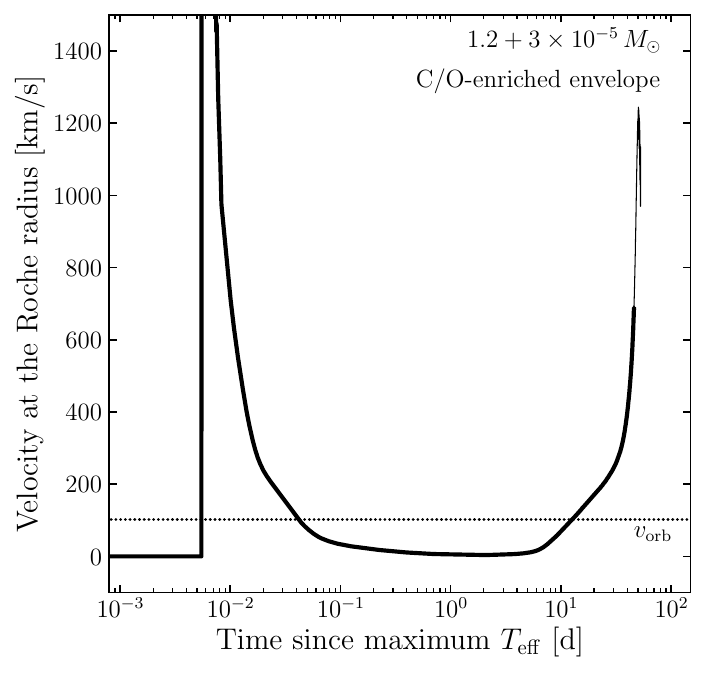}
	\caption{Time evolution of the velocity at the Roche radius for a $1.2 \msol$ WD with a C/O-enriched $3\E{-5} \msol$ outbursting envelope.  The  thin line represents the evolution after the photosphere recedes to the location of the iron opacity bump  and is only shown for completeness, as our models are not quantitatively accurate during this phase.  The Roche radius is calculated assuming a $0.3 \msol$, $0.3 \, R_\odot$ Roche-lobe-filling donor, and the dotted line is the WD's orbital velocity in such a CV system.}
	\label{fig:vvst}
\end{figure}

\begin{figure}
	\includegraphics[width=\columnwidth]{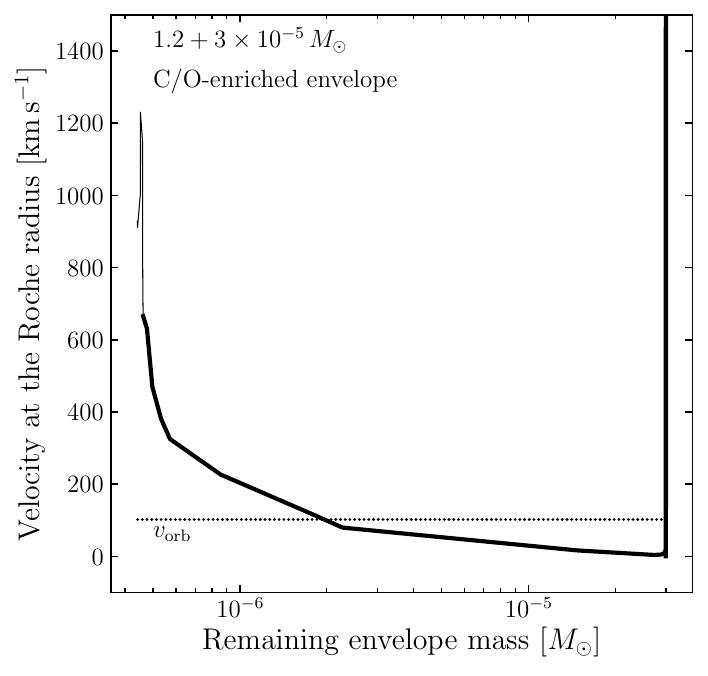}
	\caption{Same as Figure \ref{fig:vvst}, but vs.\ the remaining envelope mass, defined as the mass contained between the maximum in the energy generation rate and the $\tau=2/3$ photosphere.  Most of the mass is lost at low speeds for which the binary interaction cannot be neglected.}
	\label{fig:vvsmenv}
\end{figure}

During the phase when the critical point is outside the Roche lobe, the velocity of material at the Roche radius remains low, only accelerating after it expands beyond this location.  This can be seen in Figures \ref{fig:vvst} and \ref{fig:vvsmenv}, which show the velocity at the Roche radius, $v_{\rm Roche}$, as a function of the time since reaching the maximum $T_{\rm eff}$ and the remaining envelope mass, respectively.    The dotted lines are the WD's orbital velocity in a CV with our assumed donor.  After the initial burst of mass loss from surface $\beta$-decays, the velocity at the WD's Roche radius drops below the WD's orbital velocity, $v_{\rm orb}$, and stays there until $\unit[11]{d}$ have passed, by which point more than 90\% of the envelope mass has been ejected.  The remaining envelope mass at the time when $v_{\rm Roche}$ increases to $v_{\rm orb}$  is given in column~5 of Table~\ref{tab} for all of the models.

\begin{table*}
\begin{center}
\caption{Critical envelope masses and wind properties for the models presented in this work}
\label{tab}
\begin{tabular}{ccccccc}
WD  & C/O & Initial  & Successful  &$M_{\rm env}$ when $v_{\rm Roche}$  & $M_{\rm env}$ when $\dot{M}_{\rm Roche}$ falls & Percentage of ejected mass \\
 mass & enrichment?  & $M_{\rm env}$  & wind? &  rises above $v_{\rm orb}$\tablenotemark{a,b} & below $10^{-7} \, M_\odot \, {\rm yr^{-1}}$\tablenotemark{a} & ejected with $v_{\rm Roche} < v_{\rm orb}$\tablenotemark{c}\\
$ [M_\odot ] $ & & $[M_\odot]$ & & $[M_\odot]$ & $[M_\odot]$\\
\hline
\hline
0.6 & No & $10^{-3}$ & No & --- & $5.8\E{-5}$ & 100\% \\
0.6 & No & $3\E{-4}$ & No & --- & $5.8\E{-5}$ & 100\%\\
0.6 & No & $10^{-4}$ & No & --- & $5.9\E{-5}$ & 100\%\\
\hline
0.6 & Yes & $10^{-3}$ & Transitional & $2.1\E{-5}$ & $1.9\E{-5}$ & 100\%\\
0.6 & Yes & $3\E{-4}$ & Transitional & $2.5\E{-5}$ & $2.0\E{-5}$ & 98\%\\
0.6 & Yes & $10^{-4}$ & Transitional & $2.5\E{-5}$ & $2.1\E{-5}$ & 95\%\\
\hline
0.8 & No & $3\E{-4}$ & Transitional & $2.2\E{-5}$ & $1.4\E{-5}$ & 97\%\\
0.8 & No & $10^{-4}$ & Transitional & $2.3\E{-5}$ & $1.4\E{-5}$ & 89\%\\
0.8 & No & $3\E{-5}$ & Transitional & $3.0\E{-5}$ & $1.4\E{-5}$ & 0\%\\
\hline
0.8 & Yes & $3\E{-4}$ & Transitional & $9.2\E{-6}$ & $6.4\E{-6}$ & 99\%\\
0.8 & Yes & $10^{-4}$ & Transitional & $1.1\E{-5}$ & $6.5\E{-6}$ & 95\%\\
0.8 & Yes & $3\E{-5}$ & Yes & $1.2\E{-5}$ & $6.6\E{-6}$ & 77\%\\
0.8 & Yes & $10^{-5}$ & Yes & $1.0\E{-5}$ & $6.5\E{-6}$ & 0\%\\
\hline
1.0 & No & $10^{-4}$ & Transitional & $1.2\E{-5}$ & $4.7\E{-6}$ & 93\%\\
1.0 & No & $3\E{-5}$ & Yes & $1.3\E{-5}$ & $4.7\E{-6}$ & 67\%\\
1.0 & No & $10^{-5}$ & Yes & $1.0\E{-5}$ & $5.0\E{-6}$ & 0\%\\
\hline
1.0 & Yes & $10^{-4}$ & Yes &  $4.6\E{-6}$ & $2.2\E{-6}$ & 98\%\\
1.0 & Yes & $3\E{-5}$ & Yes &   $5.9\E{-6}$ & $2.3\E{-6}$ & 87\%\\
1.0 & Yes & $10^{-5}$ & Yes & $1.0\E{-5}$ & $2.3\E{-6}$ & 0\%\\
\hline
1.2 & No & $3\E{-5}$ & Yes &  $5.5\E{-6}$ & $1.1\E{-6}$ & 85\%\\
1.2 & No & $10^{-5}$ & Yes &   $6.0\E{-6}$ & $1.2\E{-6}$ & 45\%\\
1.2 & No & $3\E{-6}$ & Yes &  $3.0\E{-6}$ & $1.1\E{-6}$ & 0\%\\
\hline
1.2 & Yes & $3\E{-5}$ & Yes & $2.0\E{-6}$ & $4.6\E{-7}$ & 95\%\\
1.2 & Yes & $10^{-5}$ & Yes & $2.6\E{-6}$ & $4.9\E{-7}$ & 78\%\\
1.2 & Yes & $3\E{-6}$ & Yes &  $3.0\E{-6}$ & $5.2\E{-7}$ & 0\%\\
\end{tabular}
\end{center}
\tablenotetext{a}{For models with successful winds, the envelope mass is calculated as the mass between the location of the maximum burning rate and the $\tau=2/3$ photosphere.  For transitional models and models with no winds, the outer envelope mass is measured at $10 \, R_\odot$.}
\tablenotetext{b}{The brief initial phase when a small amount of mass is ejected at high velocities in some models due to $\beta$-decays near the surface is neglected in this column.}
\tablenotetext{c}{We assume that any mass that reaches the Roche radius is successfully ejected, and that mass ejection ends when $\dot{M}_{\rm Roche}$ falls below $10^{-7} \smpy$.}
\end{table*}

During this phase, the companion must play a crucial role in driving the outflow.  When $v_{\rm Roche} \ll v_{\rm orb}$, the outflow will be analogous to standard Roche lobe overflow, with material attempting to leave the WD's Roche lobe through the inner Lagrange point.  However,  the companion is overflowing its own Roche lobe and thus cannot accrete mass.  Instead, the gradually outflowing material will expand into the circumbinary environment and make its way out of the $L_2$ outer Lagrange point.  From there, the torque from the binary will send the material out into the equatorial plane as a spiral wave, which shocks upon itself and forms an outflowing equatorial torus at $\sim 10$ times the binary separation, similarly to mass outflows in non-degenerate contact binaries \citep{kuip41a,shu79a,pejc14a,pejc16a,pejc16b,pejc17a}.

For higher velocities up to $\sim  v_{\rm orb}$, the outflowing material is gravitationally focused by the companion into the equatorial plane in one or more outgoing spiral tails \citep{naga04a,jaha05a,liu17a,sala18a,sala19a,macl20a,schr21a}.  The morphology of this outflow will be somewhat different from the $v_{\rm Roche} \ll v_{\rm orb}$ case, but the important points are that the outflow is  equatorially focused, its speed is influenced by the binary interaction, and it carries away orbital angular momentum.  The critical value of $v_{\rm Roche} / v_{\rm orb}$ below which  the binary shapes the outflow depends on the mass ratio, the wind profile, and other factors.  \cite{schr21a} find a net decrease in the orbital separation due to the loss of angular momentum up to an ejection velocity of $v_{\rm orb}$ for a mass ratio of $1/3$; for larger mass ratios (i.e., lower WD masses in the context of our study), mass loss leads to orbit shrinking up to even higher velocities of several times $v_{\rm orb}$.  To be conservative, we consider mass loss to be initiated and shaped by the binary up to a velocity of $v_{\rm Roche} = v_{\rm orb}$ for all of our cases.

\begin{figure}
	\includegraphics[width=\columnwidth]{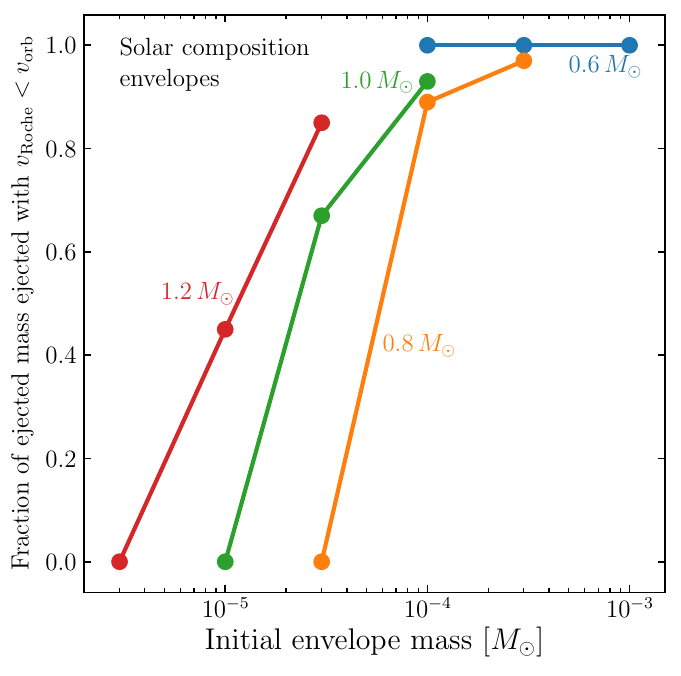}
	\caption{Fraction of the ejected mass that is ejected with $v_{\rm Roche} < v_{\rm orb}$ vs.\ initial envelope mass for the solar composition envelope models.  The WD masses are as labeled.}
	\label{fig:fracvsmenv_solar}
\end{figure}

\begin{figure}
	\includegraphics[width=\columnwidth]{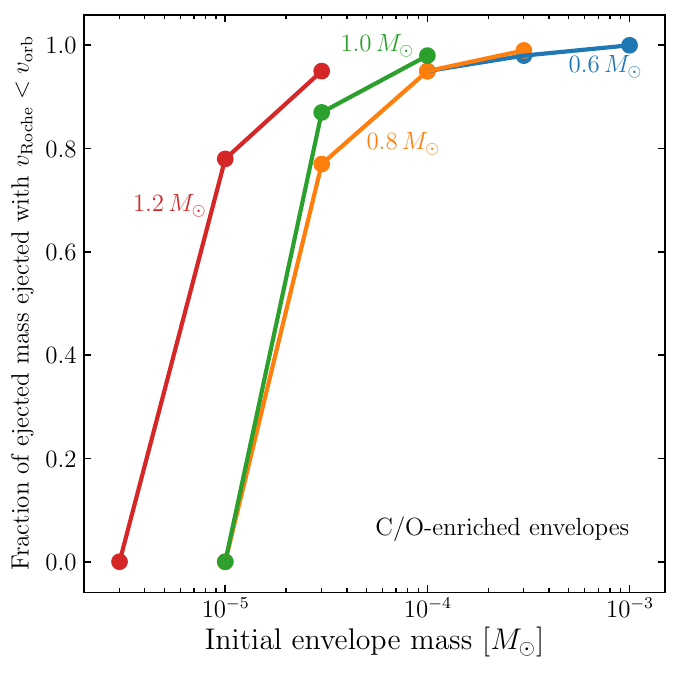}
	\caption{Same as Fig.\ \ref{fig:fracvsmenv_solar}, but for the C/O-enriched models.}
	\label{fig:fracvsmenv_co}
\end{figure}

As the nova evolves and more of the envelope is lost, the acceleration region moves inwards and the velocity of the wind continues to increase.  By day 37 for the $1.2+3\E{-5} \msol$ C/O-enriched model, the iron opacity bump peaks at the WD's Roche radius, and by day 45, the acceleration region has receded almost entirely within the Roche radius.  It is likely that, by this point, a more spherical optically thick wind can successfully be launched since the wind reaches a velocity larger than the WD's orbital velocity before crossing the Roche lobe.  However, multi-dimensional radiation hydrodynamic calculations are necessary to fully confirm this outcome.  Crucially, this only occurs once the large majority of the envelope mass (94\% for the C/O-enriched $1.2+3\E{-5} \msol$ model) has been removed; prior to this point, the binary was primarily responsible for setting the properties of the ejected mass.  This is also the case for most of the other models; see column 7 of Table \ref{tab} and Figures \ref{fig:fracvsmenv_solar} and \ref{fig:fracvsmenv_co}, which show the ratio of the mass ejected with $v_{\rm Roche}<v_{\rm orb}$ to the total mass ejected.  It is clear that  standard optically thick winds are the main mechanism of mass loss only for the models with the lowest envelope masses we consider.  These correspond to systems with accretion rates of $>10^{-8} \smpy$ \citep{yaro05,chom21a}, which  yield recurrent novae \citep{hk01,scha10a} but are a small minority of CVs \citep{pala22a}.

The $\tau=2/3$ photosphere, shown as circles in Figures \ref{fig:prof1_12_3e-5_co}, \ref{fig:prof2_12_3e-5_co}, and \ref{fig:prof3_12_3e-5_co}, continues to move inwards in radius with time, to the point that the iron opacity bump is outside the photosphere during the final phases of the simulations with successfully launched outflows.  By default, \mesa does not differentiate between radiation effects in optically thick and optically thin regions.  For instance, it is assumed that the radiation and matter are always in thermal equilibrium, so that the radiation pressure is only a function of temperature.  As a result, the velocity profile found by \mesa continues to accelerate through the iron opacity bump even when the material is optically thin.  Physically, the low optical depth should imply a low interaction probability between photons and envelope material, reducing the radiative acceleration.  However, this will be mitigated by line-broadening due to the velocity gradient, which will increase the opacity from the zero-velocity Rosseland mean value used here \citep{cast75a}.  A complete physical description of wind acceleration in this regime necessitates a hydrodynamical radiative transfer calculation to account for these effects, ideally in multiple dimensions to allow for clumping and porosity \citep{pacz69a,shav01a,owoc04a,jian15a} and including non-local thermodynamic equilibrium effects \citep{haus95a,haus96a,haus97a}.  This is outside the scope of the present work and will be the subject of future study.  

\begin{figure}
	\includegraphics[width=\columnwidth]{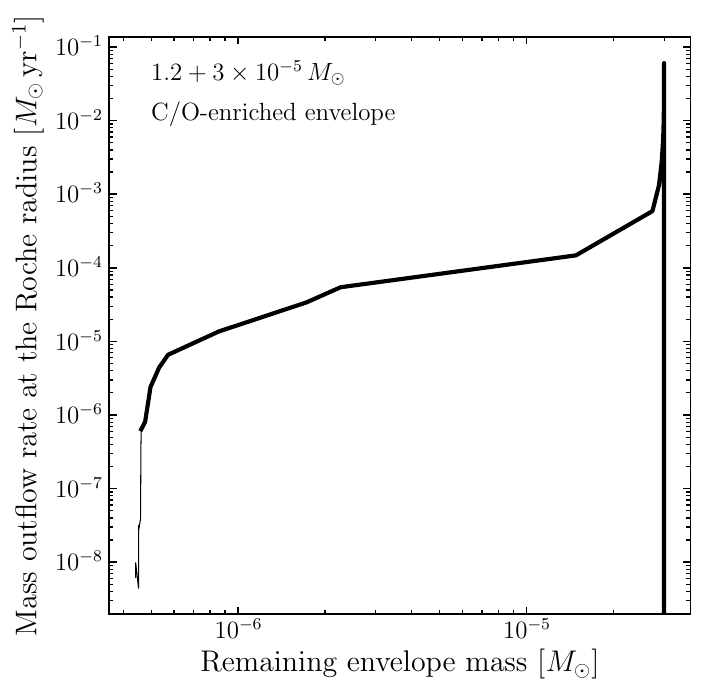}
	\caption{Mass outflow rate at the Roche radius vs.\ the remaining envelope mass for the $1.2+3\E{-5} \msol$ C/O-enriched envelope model.  The thin line represents the evolution after the photosphere recedes to the iron opacity bump  when a more accurate radiation transport calculation for the wind is needed.}
	\label{fig:mdotvsmenv}
\end{figure}

While the lack of self-consistency renders the calculations untrustworthy during this stage, it is important to note that  this phase only occurs near the end of the simulations.  Figure \ref{fig:mdotvsmenv} shows that the mass outflow rate declines precipitously as the remaining envelope mass approaches the maximum allowed mass for a steadily burning, thermally stable hydrostatic envelope, marking the end of the mass outflow stage of the nova outburst.  (The remaining envelope masses for our models when the mass outflow rate at the Roche radius, $\dot{M}_{\rm Roche}$, drops below $10^{-7} \msol \, {\rm yr^{-1}}$ are listed in column 6 of Table \ref{tab}.)  It is during this sharp drop in mass outflow rate at the end of the nova wind phase, after almost all of the mass has been ejected, that the photosphere recedes within the iron opacity bump, as shown by the thin lines in Figure \ref{fig:vvst}, \ref{fig:vvsmenv}, and \ref{fig:mdotvsmenv}.  Thus, even though the \mesa calculation is not self-consistent near the end of the outburst, most of the simulation takes place while the acceleration region is below the photosphere.  Our main conclusion that binary interaction causes  most of the mass loss  is therefore robust.

\begin{figure}
	\includegraphics[width=\columnwidth]{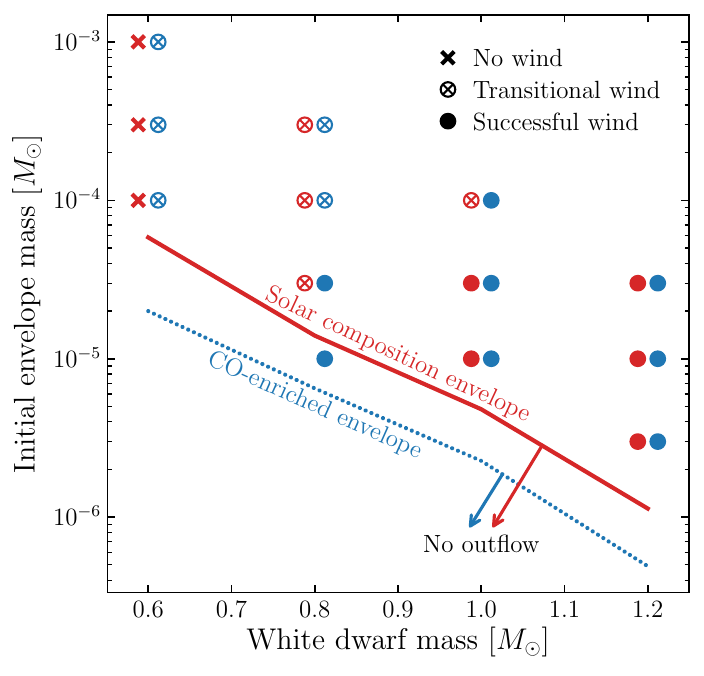}
	\caption{Outcomes of outbursts for given initial envelope mass vs.\ WD mass, as labeled.  Red points shifted slightly to the left of their actual WD masses denote models with solar composition envelopes; blue points to the right of their WD masses are models with CO-enriched envelopes.  The solid and dotted lines mark the approximate envelope mass below which thin, steady-state, thermally stable hydrostatic envelope solutions exist and outflows cease.}
	\label{fig:menvvsmwd}
\end{figure}

Simulations that launch optically thick winds that become completely unbound are shown as filled red and blue circles in Figure \ref{fig:menvvsmwd}.    The red solid and blue dotted lines show the envelope masses for the labeled compositions at which the mass outflow rates decline sharply, marking the end of the mass loss phase for models with successful winds.  Higher WD masses, lower envelope masses, and more C/O enrichment all help to launch optically thick winds in isolation.  However, not all simulations are able to successfully form winds.  We discuss these other outcomes in the following subsections.


\subsection{Models with no winds}

\begin{figure}
  \centering
  \includegraphics[width=\columnwidth]{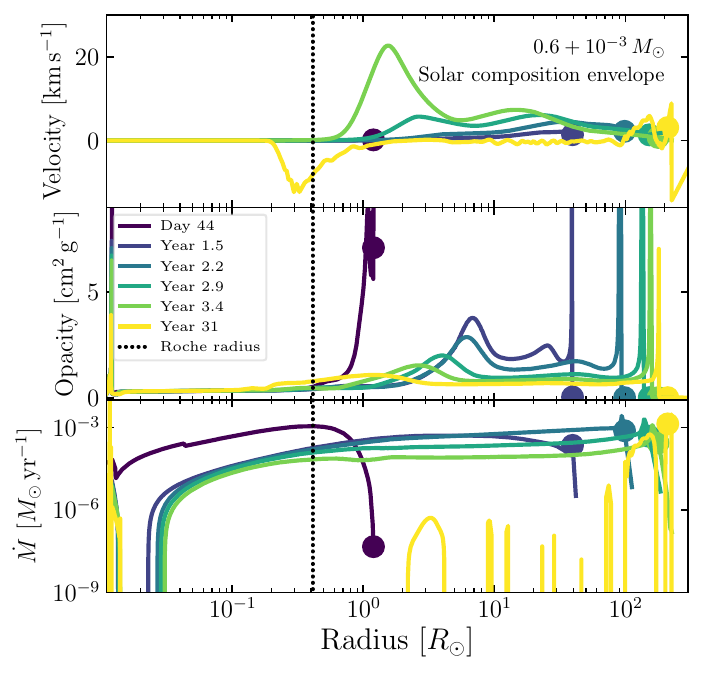}
	\caption{Same as Fig.\ \ref{fig:prof1_12_3e-5_co}, but for the $0.6 + 10^{-3} \msol$ solar composition envelope model.    Note the changes in axes and times.}
	\label{fig:prof1_06_1e-3_solar}
\end{figure}

\begin{figure}
  \centering
  \includegraphics[width=\columnwidth]{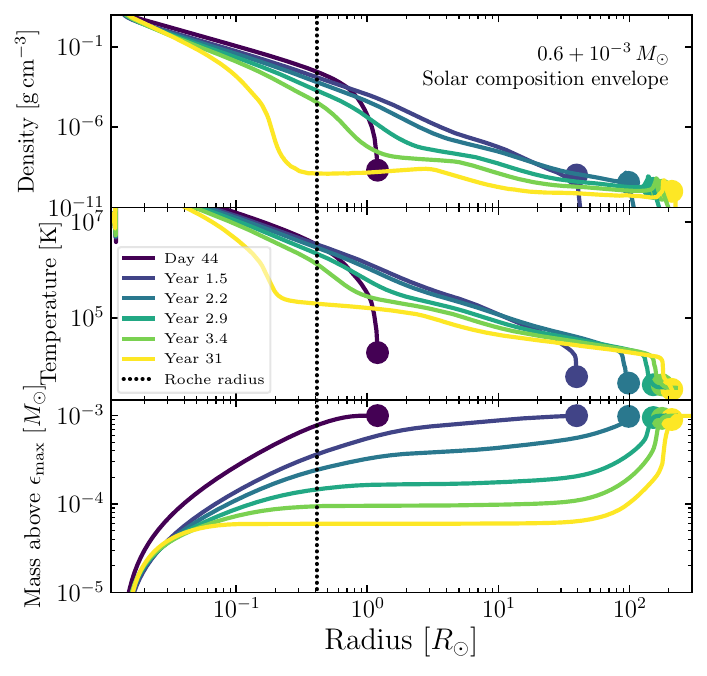}
	\caption{Same as Fig.\ \ref{fig:prof2_12_3e-5_co}, but for the $0.6 + 10^{-3} \msol$ solar composition envelope model.  Note the changes in axes and times.}
	\label{fig:prof2_06_1e-3_solar}
\end{figure}

In sharp contrast to our simulations with successful optically thick winds, the  $0.6 \msol$ models with solar composition envelopes, shown as red Xs in Figure \ref{fig:menvvsmwd}, never launch steady-state outflows.  These envelopes remain completely gravitationally bound for their entire evolution, with velocities that never approach the sound speed or the escape speed.  Radial profiles at various times during the evolution of the $0.6 + 10^{-3} \msol$ solar composition envelope simulation are shown in Figures \ref{fig:prof1_06_1e-3_solar} and \ref{fig:prof2_06_1e-3_solar}.  The lack of a steady state wind is readily apparent in the top panel of Figure \ref{fig:prof1_06_1e-3_solar}, which shows that the velocities remain in the tens of $\unit[]{km \, s^{-1}}$ within the envelope and asymptote to zero at the photosphere.  By the end of the simulation, $\unit[31]{yr}$ after the beginning of the calculation, the mass flux throughout the envelope has dropped to nearly zero  with small spikes due to weak shocks in the outer envelope.  The envelope is now in a steady state similar to a red giant branch star: an expanded, hydrostatic envelope burning hydrogen at its base, surrounding a degenerate core.  If such an outbursting WD were in a CV, the companion would necessarily drive off most of the envelope, leaving only a small remnant mass equal to the maximum mass for a thin, hydrostatic, thermally stable envelope solution.


\subsection{Transitional models}

\begin{figure}
  \centering
  \includegraphics[width=\columnwidth]{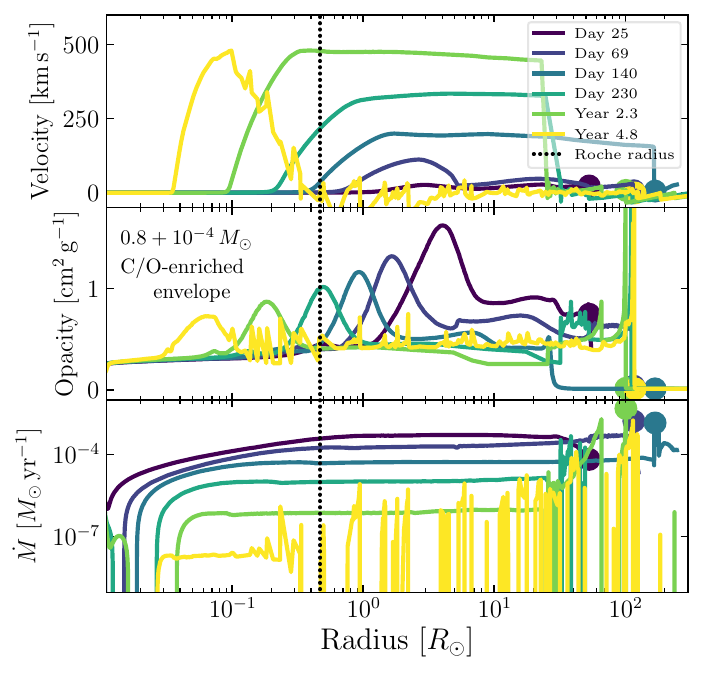}
	\caption{Same as Fig.\ \ref{fig:prof1_12_3e-5_co}, but for the $0.8 + 10^{-4} \msol$ C/O-enriched envelope model.    Note the changes in axes and times.}
	\label{fig:prof1_08_1e-4_co}
\end{figure}

\begin{figure}
  \centering
  \includegraphics[width=\columnwidth]{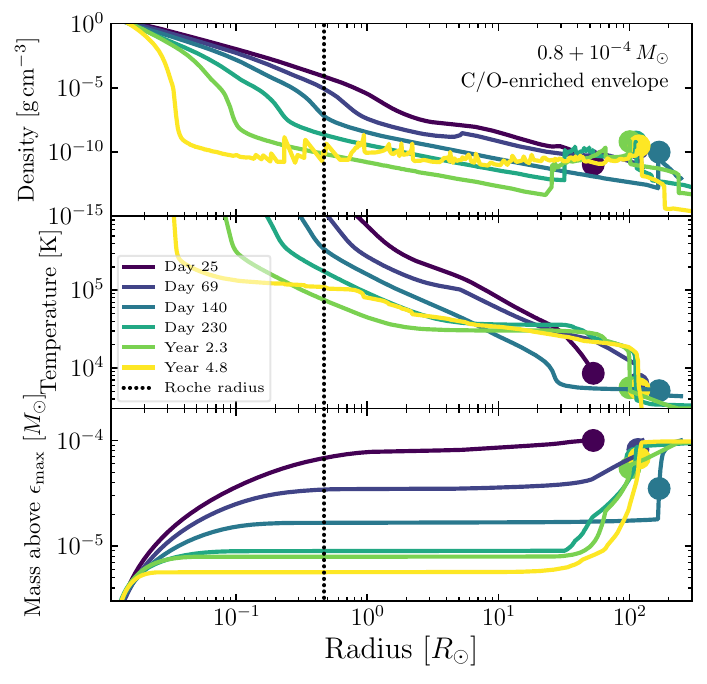}
	\caption{Same as Fig.\ \ref{fig:prof2_12_3e-5_co}, but for the $0.8 + 10^{-4} \msol$ C/O-enriched envelope model.  Note the changes in axes and times.}
	\label{fig:prof2_08_1e-4_co}
\end{figure}

Some simulations launch an optically thick wind that remains enclosed by a layer of outer material that is too massive to accelerate.  As a result, the envelope is never completely unbound: the material at large radii stops the progress of the wind.  Eventually, the wind is completely choked, and all of the material falls below the local escape speed from the WD.  We call these ``transitional'' cases, shown as combined Xs and open circles in Figure \ref{fig:menvvsmwd}.  

Figures \ref{fig:prof1_08_1e-4_co} and \ref{fig:prof2_08_1e-4_co} show radial profiles of the $0.8 + 10^{-4} \msol$, C/O-enriched envelope model, which is an example of a transitional case.  As seen in the top panel of Figure \ref{fig:prof1_08_1e-4_co}, a steady-state optically thick wind in the inner regions develops by day 140, but the mass profile shown in the bottom panel of Figure \ref{fig:prof2_08_1e-4_co} demonstrates that most of the outer envelope mass remains at low velocities.  This bound mass at the surface of the envelope blocks the outflow, preventing the formation of a successful wind.  After $\unit[3.6]{yr}$ have passed, the wind collapses into a breeze configuration \citep{park65a} similar to the profile shown at year 4.8.

It is likely that if the companion were taken into account, binary interaction would remove enough of the outer material in the early phase of the outburst to allow for the successful formation of a completely unbound wind during the later stages of mass loss.  As with the previous outcomes, it remains the case that most of the  mass ejection would be initiated by the binary interaction and not by an isotropic optically thick wind; see column~5 of Table~\ref{tab} and Figures \ref{fig:fracvsmenv_solar} and \ref{fig:fracvsmenv_co}.


\section{Comparison to previous work}
\label{sec:comp}

In this section, we compare the results of our numerical hydrodynamic calculations to previous analytic and numeric work on nova outbursts.

\cite{kh94}'s frequently cited analytic study of optically thick winds has formed the basis for our theoretical understanding of classical nova outflows for almost 30 years.  They integrate the standard wind equations (e.g., \citealt{bond52a} and \citealt{park58a,park65a}) inwards and outwards from the sonic point to generate time-independent envelope solutions in which winds are accelerated off of the iron opacity bump.  Our numerical time-dependent envelope profiles and relations such as the mass outflow rate and asymptotic velocity as a function of the envelope mass broadly agree with their analytic solutions, although quantitative comparisons are difficult due to the differences in our envelope compositions.

In their more recent work \citep{kh09}, they perform some calculations with solar composition envelopes, which we can directly compare to our results.  In their Figure 10, we see that the boundaries for envelope masses that successfully launch winds vs.\ those that instead adopt ``breeze'' solutions are shifted compared to our findings, which are summarized in Figure \ref{fig:menvvsmwd}: they find successful winds for WDs as low-mass as $\simeq 0.5 \msol$, while our solar composition simulations only launch successful winds for masses $\gtrsim 0.8 \msol$, with transitional or failed winds below this mass.

The most obvious difference between our methodologies  is their use of time-independent, steady-state solutions vs.\ our time-dependent simulations.  This is particularly obvious during the early stages of evolution, as the envelope expands and readjusts towards its steady state wind profile.  For example, the excess mass at large radii that ultimately halts the wind in our transitional models will not be captured in their time-independent calculations.  This is likely the reason for our disagreement regarding the success of winds for lower-mass WDs.

\cite{pk95} and \cite{yaro05}, and works that build off of these two studies such as \cite{epel07} and \cite{hill14a,hill16a}, present a large suite of time-dependent nova models, with a  range of accretion  rates onto WDs of a variety of masses and core temperatures.  Because our work sets the convective envelope masses by hand to avoid the complications of the pre-outburst accretion, diffusion, and mixing, none of the parameters at the onset of the outbursts described in our two sets of studies match exactly.  Still, some models provide a qualitative and informative comparison.

Each of \cite{pk95} and \cite{yaro05}'s $0.65 \msol$, $10^{-8} \smpy$ models results in a convective envelope mass of $\sim 10^{-4} \msol$ with $Z \sim 0.02$, similar to our $0.6 + 10^{-4} \msol$ solar metallicity models.  However, they find that these envelopes are ejected from the binary, albeit at low velocities $\sim \unit[130]{km \, s^{-1}}$,  whereas none of our $0.6 \msol$ models successfully launches a sustained wind.  The disagreement is even more stark for their cold (core temperature of $\unit[10^7]{K}$) $0.65 \msol$, $10^{-10}$ and $10^{-11} \smpy$ models, which yield convective envelope masses of $ 2.4-2.8\E{-4} \msol$ with $Z=0.1-0.2$, roughly similar to our C/O-enriched $0.6 + 3\E{-4} \msol$ model.  They find that these models yield  ejected envelopes with an average velocity of $\unit[600-2700]{km \, s^{-1}}$, in direct contrast with the gravitationally bound envelopes exhibited by our models.  In fact, \cite{pk95}, \cite{yaro05}, and  \cite{sib09}, which all use the same code, found mass ejection during nova outbursts for WDs as low in mass as $0.4 \msol$, in clear contrast with the present work and the analytic work of \cite{kh09}.

\cite{yaro05}'s cold $1.0 \msol$, $10^{-11} \smpy$ model results in an ejected envelope of $10^{-4} \msol$ with $Z = 0.2$, with an average ejection velocity of $\unit[1100]{km  \, s^{-1}}$, which is not too dissimilar from the outcome for our $1.0 + 10^{-4} \msol$ C/O-enriched model.  However, the timescale over which this is achieved differs drastically between our two studies: \cite{yaro05} find a mass-loss phase that lasts for $\unit[35]{d}$, whereas we find that mass loss persists for almost a factor of 10 longer.

One possibility for these discrepancies is \cite{pk95} and \cite{yaro05}'s mass loss criteria: once the velocity in a subsurface region in their simulations becomes supersonic and has a positive velocity gradient, mass is removed from the outer cell at the subsurface zone's rate of  $\dot{M} = 4 \pi r^2 v$.  However, it is not obvious that this subsurface mass flux is the correct value for the surface mass loss rate, particularly during the initial phase of the outburst, when the mass flux is not constant throughout the envelope.  Moreover, in our transitional models, an optically thick wind is launched but ultimately remains bound; \cite{pk95} and \cite{yaro05}'s mass loss algorithm may erroneously remove this mass.  Still, without an in-depth code comparison, it is difficult to pinpoint the ultimate reason(s) for the differences between our results and those of \cite{pk95} and \cite{yaro05}; such a comparison would be a useful future endeavor.

The \texttt{NOVA} code has a long and rich history of studies of classical nova outbursts (e.g., \citealt{kutt72a,poli95a,star98,star00,star09,star16a}).  In the most recent work to use the \texttt{NOVA} code, \cite{star20a} perform a suite of simulations of WDs of a range of masses accreting material with a variety of accretion rates and compositions.  The best matches with our models are for their simulations with solar composition envelopes and 25\% WD, 75\% solar composition envelopes.  In direct contrast with our results, and those of \cite{kh94}, \cite{pk95}, \cite{yaro05}, and others, \cite{star20a} find that very little mass is ejected for these compositions.  Essentially no mass is ejected for their solar composition envelopes, and $\leq 5\%$ of the nova envelope  is ejected for their 25\% WD, 75\% solar composition envelopes for WD masses $\leq 1.0 \msol$.  Their higher mass models do ejected tens of percent of their envelopes, but these ejected masses are still much less than the 98\% ejected fraction that we find for our C/O-enriched, $1.2 + 3\E{-5} \msol$ model, which provides the closest match.

The reason for their low ejected masses likely lies in the stopping criterion employed by \cite{star20a}.  Once mass expands to radii of a few times $\unit[10^{12}]{cm}$, their simulations are ended due to the material reaching densities below the lower bounds of their equation of state and opacity tables.  However, this condition is reached very early in the evolution of our simulations.  If we were to employ a similar condition, we would also find very low ejected masses simply because most of the mass ejection occurs after the outermost zones reach these radii.  This extended material can also fall off of our equation of state and opacity tables, but since the material is at extremely low optical depths, any uncertainties are inconsequential for the evolution of the bulk of the envelope.  Thus, we regard \cite{star20a}'s findings of very low ejected masses to be a consequence of premature stopping conditions, and we encourage them to extend their simulations to later times.

Our study is certainly not the first theoretical work to examine  the role that a companion may play in driving mass loss during novae (e.g., \citealt{macd80,macd85a,macd86a,kh94,kato11a}).  However, these previous studies only consider the companion's impact on one-dimensional spherically symmetric steady-state envelope solutions.  They do not allow for the possibility that the companion entirely disrupts the standard spherical wind solution and instead shapes the mass loss into an equatorial outflow.  We have shown that the latter case is more likely at most times for most nova systems.

\cite{livi90} infer that the donor star is important for shaping, and perhaps even driving, mass loss in classical novae due to the empirical observation that the pseudo-photosphere is located at a radius well outside of the binary separation.  However, even given this hierarchy of radii, the companion would still not dominate the outflow properties if the outflow were ejected from the WD's surface at a velocity much faster than the orbital velocity.  In the present study, we show that the outflow velocities at the Roche radius are in fact slower than the orbital velocity during the majority of the mass loss phase for most novae, and thus the companion does play a dominant role in driving mass loss during outbursts.

\cite{livi90} also perform a two-dimensional axisymmetric hydrodynamic simulation of a slow WD wind that is focused onto the equatorial plane by the companion, which is approximated as a smeared-out axisymmetric ring.  In reality, the three-dimensional configuration of the CV will instead likely result in an ejected spiral that self-shocks and develops into an outflow driven by the binary torque \citep{pejc16a,pejc16b,schr21a}.


\section{Conclusions}
\label{sec:conc}

In this paper, we have described our suite of  hydrodynamic stellar evolution simulations of isolated WDs undergoing classical nova eruptions.  We have shown that, for typical systems, most of the mass that is lost in optically thick winds reaches the WD's Roche radius at velocities lower than the orbital velocity.  As a result, when the binary companion is taken into account, the properties of mass driven from the system will actually be primarily set by the binary interaction.  This is even more true for lower-mass WDs $\lesssim 0.8 \msol$, for which no optically thick winds are successfully launched in isolation; for these systems, mass loss would not occur at all if not for the binary interaction.  Any study aiming to predict classical nova observables, from optical light curves and spectra to non-thermal shock signatures, needs to account for the dominating influence of the binary companion on the outflows.   The only CVs for which optically thick winds drive the majority of mass loss are the rare systems with high accretion rates $>10^{-8} \smpy$, which yield the lowest envelope ignition masses.

Our results predict that much of the outflowing mass in classical novae is likely to be focused into the equatorial plane in the form of spiral arms that may undergo internal shocks and dissipate and radiate some fraction of their kinetic energy,   as seen in analogous simulations of mass loss in contact binaries and high-mass X-ray binaries \citep{pejc16a,pejc16b,schr21a}.  However, these studies differ in important ways from the classical nova case, so quantitative predictions of the geometry and observables associated with this binary-induced mass loss await future dedicated multi-dimensional radiation hydrodynamical simulations.

The self-collision of the spiral arms, together with the subsequent shocks between late-time high-speed radiation-pressure-driven winds and earlier-time slower binary-driven mass loss, are plausibly responsible for the rich and complex light curves of classical novae, including the evidence for shock-generated hard X-rays and non-thermal gamma-rays (e.g., \citealt{chom21a}).  The key role of the binary in driving nova winds might also explain the super-Eddington luminosities observed for some classical novae.  These super-Eddington luminosities cannot be understood using radiation-pressure-driven winds alone but may be explained by the additional kinetic energy gained through interaction with the binary.

For  novae in symbiotic systems (e.g., \citealt{miko10a}), which have giant companions and thus much larger binary separations, the acceleration of the outflows from the WD will not be affected by the binary interaction.  However, even in these systems, the influence of the companion on the nova observables cannot be neglected.  The companion's wind will both reprocess the resulting radiation as well as lead to shock interaction with the nova outflow.

The orbital energy and angular momentum carried away by the binary-driven mass loss will have a profound effect on the long-term evolution of the CV, possibly destabilizing the binary enough to cause a merger \citep{shen15a,metz21a}.  This is especially true for CVs with lower-mass WDs, since the mass ratio is higher and because the nova ignition masses are larger, which means each nova event has more of an effect on the post-nova binary parameters.  This preferential destruction of CVs with lower-mass WDs would explain the large discrepancy between the CV mass distribution and that of field WDs and pre-CVs \citep{zoro11a,nele16a,schr16a,pala20a,pala22a} and will be the subject of  future studies.

Future work will also include multi-dimensional radiation hydrodynamics calculations to better quantify the transition between a binary-dominated outflow and the successful launching of a more spherical optically thick wind as well as hydrodynamic simulations of the interaction between these two modes of mass ejection.  Such studies will be necessary for a complete understanding of the surprisingly complex physics of classical novae and will have a broad applicability to investigations of mass loss in other contexts, such as contact binaries, common envelope systems, and high-mass X-ray binaries.


\software{\texttt{matplotlib} \citep{hunt07a}, \mesa \citep{paxt11,paxt13,paxt15a,paxt18a}}


\acknowledgments

We thank Laura Chomiuk, Dan Kasen, Brian Metzger, Raffaella Margutti, Alison Miller, Joe Patterson, Ond\v{r}ej Pejcha, Brad Schaefer, and Jeno Sokoloski for helpful discussions, Bill Paxton and the MESA Council for developing \mesa, and the referee for their thoughtful review.  This work was supported by NASA through the Astrophysics Theory Program (80NSSC20K0544) and by the Research Corporation for Science Advancement through a ``Scialog: Time Domain Astrophysics'' award.  This research used the Savio computational cluster resource provided by the Berkeley Research Computing program at the University of California, Berkeley (supported by the UC Berkeley Chancellor, Vice Chancellor of Research, and Office of the CIO).


\appendix

The hydrodynamic \mesa calculations presented in this work are informed by previous studies that used \mesa to simulate hydrodynamical outflows \citep{quat16a,full17a,shen17a}.  These studies helped to define the necessary flags and parameters to set in the inlist, which we display below, to allow for the development of outflowing winds.  

In addition to these flags, we also implement several additional modifications for our particular set of simulations.  In the initial stages of the outburst as the envelope is being accelerated to a steady wind solution, the outermost zone can have issues converging to a solution.  To circumvent this roadblock, we restrict each cell's mass to be at least  $10^{-12}$ of the enclosed mass at that coordinate until the photosphere reaches $1  \, R_\odot$.  After this point, the minimum cell mass is reduced to $10^{-99} $ of the enclosed mass at that coordinate; i.e., cells are allowed to become as small as necessary, ensuring the simulation is not starved for gridpoints.

Once the outflow has begun, an opacity minimum of $\unit[0.01]{cm^2 \, g^{-1}}$ is established in the optically thin region above the photosphere to increase numerical stability.  For some models that encounter numerical difficulties, the opacity in this region is later further constrained to be a constant $\unit[0.01]{cm^2 \, g^{-1}}$.

\

{\ttfamily
\noindent \&star\_job

\

      ! initialize the model

      load\_saved\_model = .true.

      saved\_model\_name = '1.0\_2e7\_1e-4\_0.2.mod'

      set\_initial\_age = .true.

      initial\_age = 0

      set\_initial\_model\_number = .true.

      initial\_model\_number = 0

\

      ! set the outer optical depth to be $0.01 \times 2/3$

      set\_to\_this\_tau\_factor = 1e-2

      set\_tau\_factor = .true.

\

      ! set the nuclear network

      change\_net = .true.

      new\_net\_name = 'cno\_extras.net'

\

      ! turn on hydrodynamics

      change\_v\_flag = .true.

      new\_v\_flag = .true.

\

\noindent / ! end of star\_job namelist

\

\

\noindent \&controls

\

      min\_dq = 1d-99 ! 1d-12 until photosphere reaches 1 Rsol, then switch to 1d-99

      logQ\_min\_limit = -99

      min\_timestep\_limit = 1d-99

        use\_Ledoux\_criterion = .true.

\

      ! set outer boundary conditions for hydrodynamics

      use\_momentum\_outer\_BC = .true.

      use\_compression\_outer\_BC = .true.

      use\_zero\_Pgas\_outer\_BC = .true.

\

      ! limit convective velocity and acceleration

      max\_conv\_vel\_div\_csound = 1d0

      mlt\_accel\_g\_theta = 1

      min\_T\_for\_acceleration\_limited\_conv\_velocity = 0

\

      ! turn on artificial viscosity

      use\_artificial\_viscosity = .true.

\

      ! use Type 2 opacities to account for C/O enrichment
      
      use\_Type2\_opacities = .true.

      Zbase = 0.02

\

        merge\_if\_dlnR\_too\_small = .true. ! turned on near the end of some runs to stabilize shocks

        mesh\_min\_dlnR = 1d-7

\

      use\_other\_kap = .true. ! set an opacity floor for optically thin material

\

\noindent / ! end of controls namelist
}



\end{document}